\documentstyle[12pt]{article}
\newcommand{\T}{\mbox{\rm Tr}}
\newcommand{\MPL}{Mod. Phys. Lett. {\bf A}}
\newcommand{\NPB}{Nucl. Phys. {\bf B}}

\newcommand{\PLB}{Phys. Lett. {\bf B}}
\newcommand{\IJM}{Int. Jour. Mod. Phys. {\bf A}}

\begin{document}

\begin{flushright}
hep-th/9804206 \\
BROWN-HET-1114  \\
March 1998
\end{flushright}     
\begin{center}
{\Large 
{\bf Large $N$ Field Theory Of  ${\mathbf {\cal N}=2}$ Strings and Self-Dual
Gravity
\footnote{
supported by 
DOE under grant DE-FG0291ER40688-Task A.\\
email address: antal,nunes@het.brown.edu, mm@barus.physics.brown.edu}}}

Antal Jevick${\rm i}^{\star}$, Mihail Mihailesc${\rm u}^{\star}$ 
and Jo\~ao P. ${\rm Nunes}^{\star ,\dagger}$\\

Department of Physic${\rm s^{\star}}$\\
Brown University\\
Providence RI 02912, USA

and

Departamento de Matem\'ati${\rm ca^{\dagger}}$\\
Instituto Superior T\'ecnico\\
Av. Rovisco Pais, 1096 Lisboa Codex, Portugal

{\bf Abstract}
\end{center}

We review some aspects of the construction of self-dual gravity and the associated 
field theory of ${\cal N}=2$ strings in terms of two-dimensional sigma models at 
large $N$. The theory is defined through a large $N$ Wess-Zumino-Witten model in a 
nontrivial background and in a particular double scaling limit.
We examine the canonical structure of the theory and describe an 
infinite-dimensional Poisson algebra of currents.

\vspace{.7cm}
(Review to appear in the special issue of the ``Journal of Chaos, Solitons and 
Fractals'')
\newpage

\section{Introduction}

String theories exhibit remarkable physical and mathematical properties and their 
description in terms of 
fields in the target space-time where the string propagates is interesting and relevant 
for both the study of 
non-perturbative aspects of the theory and for the understanding of its 
fundamental meaning .
Of recent major interest is the nature and origin of gravitational interactions 
in string theory and its possible partonic constituent structure.
The ${\cal N}=2$ string\footnote{For an excelent review see \cite{M}.} 
\cite{dadda, dadda1} is one of the simplest string theories and posesses 
remarkable properties. In the case of the closed string the spectrum consists 
essentially of one real massless scalar field which for the critical 
theory lives in a target space-time that is a two-dimensional complex manifold 
of signature $(2,2)$ and equipped with a (pseudo) Kahler metric \cite{dadda}-\cite{L}. 
This theory shares some similarities with the two-dimensional theory of non-critical 
strings where the only bulk degree of freedom was a masless scalar
\cite{KLE}. 
In the present case the scalar field $\Omega = \Omega(y,{\bar y},z,{\bar z})$ 
parametrizes possible metric deformations of the free ${\cal N}=2$ supersymmetric 
sigma model action
$$
S_{0}=\int d^{2}xd^{2}\theta d^{2}{\bar \theta} K_{0}(y,{\bar y},z,{\bar z})
$$
where $y$ and $z$ are complex coordinates on the target space and 
where the flat Kahler potential is given by $K_{0}=y{\bar y}-z{\bar z}$. One has the
deformations  $K_{0}\rightarrow K=K_{0}+\Omega$ describing a non-flat Kahler metric 
$g_{i{\bar j}}=\partial_{i}{\bar \partial_{j}}K$ \cite{OV}. This geometrical 
interpretation of the field $\Omega$ follows from an examination of its 
equation of motion which is determined by the string theory amplitudes. Due to the 
strong kinematical constraints that exist in a space-time of signature $(2,2)$ the 
${\cal N}=2$ string amplitudes have peculiar properties which make them particularly 
simple. In fact, all tree level amplitudes beyond and including the 4-point function 
vanish and this result is expected to hold at loop level as well. This is
an indication of some underlying topological content of the theory. The tree level 
3-point string amplitude, as well as the vanishing of the 4-point and higher tree
level correlation functions as demonstrated by Ooguri and Vafa, is reproduced by the 
cubic action for $\Omega$ \cite{OV}
\begin{equation}
S_{cubic}=\int\;d^{2}yd^{2}z \exp(\frac{1}{2}(\partial_{y}\Omega\partial_{\bar y}\Omega
-\partial_{z}\Omega\partial_{\bar z}\Omega)+\frac{1}{3}\Omega\{\partial_{y}\Omega,
\partial_{z}\Omega\}),
\label{1.1}
\end{equation}
which gives as equation of motion the Pleba\'nski first heavenly equation for self-dual 
gravity \cite{Pl}
\begin{equation}
\det(g_{i{\bar j}})\;=\;-1\;\;\;\longleftrightarrow\;\;\;\partial^{2}_{y{\bar y}}\Omega -
\partial^{2}_{z{\bar z}}\Omega -
\{\partial_{y}\Omega ,\partial_{z}\Omega\}=0,
\label{1.2}
\end{equation}
where $\{\; ,\;\}$ denotes the Poisson bracket 
$\partial_{\bar y}\partial_{\bar z}-\partial_{\bar z}\partial_{\bar y}$ \cite{P}. 
Since the determinant of the Kahler 
metric is constant, the target space is a Ricci flat four-dimensional Kahler manifold,
and therefore it is also hyperkahler and the Riemann tensor is self-dual (or 
anti-self-dual). 
Thus, as we should expect of a theory of closed strings, the closed ${\cal N}=2$ 
string also contains gravity and we may consider it as a definition of what a quantum 
theory of self-dual gravity should be.
A proper understanding of the theory would also have impications to other subjects of 
recent interest, for example the quantum theory of membranes \cite{J,KM,G}.

Although the cubic action in (\ref{1.1}) describes the string tree level amplitudes 
quite well, at one-loop level some problems arise \cite{MM,BGI,M}. For example, the string 
one-loop partition function behaves as $\sim \tau_{2}^{-1}$ for small $\tau_{2}$ where 
$\tau_{2}$ is the imaginary 
part of the modular parameter describing the moduli space of tori, while the field theory 
one-loop 
partition function for a massless scalar in four real dimensions would be expected to 
behave as
$\sim s^{-2}$ for small $s$ where $s$ is a Schwinger parameter which is the 
field theory analog of $\tau_{2}$. 
The 
behaviour of the string one-loop partition function corresponds instead to a massless 
scalar 
field living in {\it two} real dimensions. A similar mismatch occurs for the one-loop 
3-point 
function where the string gives an infrared divergence of the form $\sim\tau_{2}^{2}$ 
for large 
$\tau_{2}$, which would again correspond to a two-dimensional field theory, while on the 
other 
hand the cubic action (\ref{1.1}) gives instead a divergence of type $\sim s$.
Therefore, although the critical target space dimension is four real dimensions it looks 
like that 
at the quantum level the string would ``prefer'' to live in only two real dimensions
\cite{MM,BGI,M} !

For the open string the analogous situation arises where now self-dual gravity is 
replaced by 
self-dual Yang-Mills theory and the massless scalar field takes values 
in the Lie algebra of the gauge group $G$ (which for this type of strings can be chosen 
arbitrarily) corresponding to the Chan-Paton factors attached to the ends of the strings.

The problem of trying to match the amplitudes in the field theory and in the 
string theory at the quantum level, 
indicates that  some type of two-dimensional description would provide a better 
formulation of the field theory of ${\cal N}=2$ strings. Self-dual gravity as shown at 
classical level by Q.Park can indeed be interpreted as a two-dimensional sigma model 
at large $N$ . The basic idea \cite{P} is to use the fact that as $N\rightarrow
\infty$ the
Lie algebra of $SU(N)$ ``approaches'' the Poisson algebra of functions on some 
two-dimensional 
surface $\Sigma$ with the Lie bracket at finite $N$ being replaced by the Poisson bracket 
when $N\rightarrow\infty$ \footnote{One should be 
careful about the way in which the limit is taken as several different 
large $N$ limits are possible. See for example \cite{BH}.} \cite{Ho}.
One starts with a two-dimensional equation of motion for some Lie algebra valued fields 
and when the Lie bracket is replaced by the Poisson bracket one effectively generates an 
equation of motion in a four-dimensional space-time with the two extra variables coming 
from the large $N$ color following the original proposal of \cite{FIT} made in the context 
of Yang-Mills theory. 

In \cite{JMN} (hereafter refered to as (I)) we have explored the approach of \cite{P} 
and considered a large $N$ Wess-Zumino-Witten model 
to establish a  field theory of ${\cal N}=2$ strings, $i.e.$ self-dual gravity. 
We have introduced a nonlinear formulation of the theory which when taken in
a particular classical background and a in a certain double scaling limit was
seen to give self-dual gravity. We have studied the corresponding scattering amplitudes 
up to one-loop level.  In this paper we review and continue to further examine the model 
formulated in (I). In section 2 we discuss some general questions on the formulation 
of self-dual gravity as a large $N$ sigma model and in section 3 we describe our 
construction. Section 4 explains the relation of the model with self-dual Yang-Mills 
theory and in section 5 we briefly comment on the field theory amplitudes obtained from 
our model. In Section 6 we establish the Hamiltonian structure of the theory and 
construct an infinite-dimensional Poisson algebra of currents.
We close in section 7 with some conclusions.


\section{Self-Dual Gravity as a Large $N$ 2D Sigma Model}

We will consider a family of two-dimensional sigma models defined in a space-time with 
Lorentzian signature ${\rm (1,1)}$, light-cone coordinates 
$x^{+}=x^{1}+x^{2}$, $x^{-}=x^{1}-x^{2}$ and metric 
$ds^{2}_{2d}=dx^{+}dx^{-}$. 
The models will consist of a  chiral model term together with a Wess-Zumino term,
where the coefficient of the chiral term will be perturbed away from the
conformal fixed point value, so that the models can be seen as  perturbed 
WZW models at level $K$,
\begin{equation}
S(g)=\frac{(1+\epsilon)K}{4\pi}\int dx^{+}dx^{-}\T
(\partial_{+}\,g\partial_{-}g^{-1}) + \frac{K}{12\pi} \Gamma_{WZW}(g),
\label{2.2}
\end{equation}
where $g$ is the group valued field,
$\epsilon$ is a real parameter and $\Gamma(g)$ is the Wess-Zumino term.
When $\epsilon =0$ we obtain the usual WZW model fixed point. The group $G$
remains unspecified for now, but shortly we will take the Lie algebra to be 
some large $N$ limit of $su(N)$.

The classical equations of motion will be given by
\begin{equation}
\frac{\epsilon}{1+\epsilon}\;\partial_{+}(g^{-1}\partial_{-}g)+
\frac{2+\epsilon}{1+\epsilon}\;\partial_{-}(g^{-1}\partial_{+}g)=0.
\label{2.3}
\end{equation}

Let us consider the case where there's only the WZW term, that is the limit 
$\epsilon\rightarrow -1$.
The equation of motion becomes,
\begin{equation}
\partial_{+}(I_{-})-\partial_{-}(I_{+})=0.
\label{new.1}
\end{equation}
where the Lie algebra valued currents are given by $I_{\pm}=g^{-1}\partial_{\pm}g$.
Of course the currents also satisfy the flatness condition,
\begin{equation}
\partial_{+}(I_{-})-\partial_{-}(I_{+})+[I_{+},I_{-}]=0.
\label{new.2}
\end{equation}
One now takes the large $N$ limit and replaces the Lie bracket $[\;,\;]$ by the 
Poisson bracket 
$\{\;,\;\}=\partial_{q}\partial_{p}
-\partial_{p}\partial_{q}$ of functions of $q$ and $p$ where $q,p$ are coordinates on 
some two-dimensional surface $\Sigma$, and one solves 
the equation of motion (\ref{new.1}) for the currents by setting 
$I_{+}=\partial_{+}\Omega$ and $I_{-}=\partial_{-}\Omega$ in terms of a scalar field 
$\Omega(x^{+},q,x^{-},p)$. 
Then the flatness condition 
(\ref{new.2}) becomes an equation of motion for $\Omega(x^{+},q,x^{-},p)$,
\begin{equation}
\{\partial_{+}\Omega,\partial_{-}\Omega\}=0.
\label{new.3}
\end{equation}
On the other hand, if one now sets  $\Omega \rightarrow\Omega -x^{+}q+x^{-}p$ in 
the first heavenly equation (\ref{1.2}) and let 
$(x^{+},q,x^{-},p) \leftrightarrow (y,{\bar y},z,{\bar z})$ one obtains
$\{\partial_{+}\Omega,\partial_{-}\Omega\}=1$. This is essentially the same as equation 
(\ref{new.3}) if one replaces the functions $\Omega(x^{+},q,x^{-},p)$ by the corresponding 
symplectic vector fields 
$\xi_{\Omega}=\partial_{q}\Omega\;\partial_{p} -\partial_{p}\Omega\;
\partial_{q}$ such that 
the Lie bracket of two such vector fields satisfies $\xi_{\{f,g\}}=[\xi_{f},\xi_{g}]$ 
and such that constant functions give the vector field zero 
\cite{P}. Therefore, classical self-dual gravity is obtained as the large $N$ limit 
of a two-dimensional sigma model.
The equation of motion (\ref{1.2}) for $\Omega$ can now be obtained from the cubic 
action (\ref{1.1}) of the previous section. However, at the quantum level we may expect 
that the large $N$ two-dimensional sigma model and the cubic action, which is a 
``pseudo-dual'' version of the original sigma model  
behave differently \cite{FJ}. Indeed, although the family of two-dimensional sigma models 
(\ref{2.2}) has a well known infinite-dimensional Poisson algebra of currents, that 
desirable feature is far from obvious in the canonical structure of the cubic action 
(\ref{1.1}). If we canonically quantized (\ref{1.1}), in the usual time-like 
quantization, we would obtain the 
momenta $\Pi = \partial_{0}\Omega$ and we would set $\{\Pi,\Pi\}_{P.B.}=0=
\{\partial_{0}\Omega,\partial_{0}\Omega\}_{P.B.}$ while we know that the 
currents $I_{0}=\partial_{0}\Omega$ in
the two-dimensional sigma model generate a rich Kac-Moody type Poisson algebra with the 
Poisson bracket given schematically by
$\{I_{0}^{\alpha},I_{0}^{\beta}\}_{P.B.}=f^{\alpha\beta}_{\gamma}I_{0}^{\gamma}$
where the $f^{\alpha\beta}_{\gamma}$ are the Lie algebra structure constants. 
It is clear that the canonical structure of the
pseudo-dual cubic action is different from the canonical structure of the original 
two-dimensional sigma model. This is unfortunate since we certainly would like to 
keep the original current algebras of the two-dimensional sigma models in our string 
field theory.

One  would therefore like to define a field theory of 
${\cal N}=2$ strings formulated as a two-dimensional nonlinear sigma model in
such a form which preserves its canonical structure. In addition, we would prefer to 
work with functions of $q$ and $p$ rather than with 
symplectic vector fields. To achieve this and to generate the four-dimensional 
space-time out of
the two dimensions $x^{+},x^{-}$ together with the large $N$ ``color'' gauge 
algebra variables $q,p$, we went back to the nonlinear version of models in (\ref{2.2}).


\section{Large $N$ WZW Field Theory of ${\mathbf {\cal N}=2}$ Strings}

Let us consider the general family of two-dimensional sigma models in (\ref{2.2}).
As long as $\epsilon\neq -2,-1,0$ in (\ref{2.3}), one can define $Q$ by
$$
\frac{2+\epsilon}{1+\epsilon}\;g^{-1}\partial_{+}g=Q^{-1}\partial_{+}Q
\;\;\;\; {\rm and} \;\;\;\;
\frac{\epsilon}{1+\epsilon}\;g^{-1}\partial_{-}g=Q^{-1}\partial_{-}Q
$$
where $Q$ is a classical solution of the pure chiral model equation obtained by 
setting $\epsilon\rightarrow\infty$, see for example\footnote{One can easily check that 
this is consistent with the flatness conditions for both the $g$ and $Q$ currents.}
\cite{Z},
$$
\partial_{+}(Q^{-1}\partial_{-}Q)+\partial_{-}(Q^{-1}\partial_{+}Q)=0.
$$
To make the connection between the two-dimensional sigma model (\ref{2.2})
and four-dimensional self-dual gravity, we start by
expanding the field $g$ aroung a specific classical configuration. For reasons
that will become clear shortly, this classical background will be given by
\begin{equation}
Q^{-1}\partial_{+}Q=\frac{{\hat q}}{\Lambda}+
\frac{x^{-}}{2\Lambda^{2}};\;\;\;\;\;\; 
Q^{-1}\partial_{-}Q=\frac{{\hat p}}{\Lambda}-
\frac{x^{+}}{2\Lambda^{2}},
\label{2.4}
\end{equation}
where the (infinite) matrices ${\hat q}$ and ${\hat p}$ satisfy 
$[{\hat q},{\hat p}] =1$ and $\Lambda$ is an arbitrary parameter.
In the context of our approach, what is important about ${\hat q}$ and 
${\hat p}$ is that in the large $N$ limit they become canonical variables
$q$ and $p$, possibly after some rescaling by factors of $N$.
The corresponding $g$ will be given by
\begin{equation}
g_{0}^{-1}\partial_{+}g_{0}=\frac{1+\epsilon}{2+\epsilon}\,
(\frac{{\hat q}}{\Lambda}+\frac{x^{-}}{2\Lambda^{2}});
\;\;\;\;\;\; 
g_{0}^{-1}\partial_{-}g_{0}=\frac{1+\epsilon}{\epsilon}\,
(\frac{{\hat p}}{\Lambda}-\frac{x^{+}}{2\Lambda^{2}}).
\label{2.5}
\end{equation}

We will now expand the action around this classical background. We take 
$N\rightarrow\infty$
such that the fields will take values in the infinite dimensional Poisson 
algebra of functions on the surface $\Sigma$, $sdiff(\Sigma )$. 
The trace becomes the integration over the $su(\infty)$ ``color'' variables $q$ and $p$.
The quantum field fluctuations will be 
described by a field $\omega$, which will take values in the Poisson 
algebra, where we define 
\begin{equation}
g=g_{0}\exp (\omega).
\label{2.6}
\end{equation}

From the Polyakov-Wigman 
formula \cite{PW,LS} and after rescaling the field $\omega$ to get a normalized  
kinetic term we obtain\footnote{The $x^{+}$ and $x^{-}$ dependent terms in 
the classical background currents (\ref{2.5}) give vanishing contributions 
to $S(\omega)$.}
\begin{eqnarray}
\nonumber
S(\omega )=\frac{1}{2}\int dx^{+}dx^{-}\T (
\partial_{p}\omega\partial_{-}\omega -\partial_{q}
\omega\partial_{+}\omega) 
\phantom{777777777777777777777777777}
\\
\nonumber
+\frac{1}{3!} g_{st}\int dx^{+}dx^{-} 
\T (\omega\{\partial_{p}\omega,\partial_{-}\omega\}
-\omega\{\partial_{q}\omega , \partial_{+}\omega\} ) 
\phantom{777777777777777}
\\
\nonumber
+\frac{1}{4!}g_{st}^{2}\int dx^{+}dx^{-} \T 
(\partial_{q}\omega\{\{\partial_{+}\omega ,\omega\} ,\omega\}
-\partial_{p}\omega\{\{\partial_{-}\omega,\omega\} ,\omega\})+\cdots
\phantom{7777777}
\\
+\Lambda\int dx^{+}dx^{-}\T (-\partial_{+}\omega\partial_{-}\omega
-\frac{2}{3!}g_{st}\omega\{\partial_{+}\omega ,\partial_{-}\omega\}
+ \frac{2}{4!}g_{st}^{2}
\omega\{\{\partial_{-}\omega ,\omega\},\partial_{+}\omega\}) + \cdots
\label{2.7}
\end{eqnarray}
where the coupling constant 
$g_{st} = \sqrt{\frac{4\pi\Lambda }{K(1+\epsilon )}}$ will be identified with the 
string coupling constant when we look at the string amplitudes as in \cite{OV}.
We see that expanding around the background generates cubic, quartic and an 
infinite number of higher point vertices for the 
field $\omega$, and this is to be compared with the purely cubic action 
(\ref{1.1}). Moreover, the two-dimensional propagator 
$\partial_{+}\omega\,\partial_{-}\omega$, together with other terms without
derivatives in the ``color directions'', is multiplied by factor of $\Lambda$.
In the limit $\Lambda\rightarrow 0$ these terms disappear and the remaining
quadratic terms in $\omega$ define 
a four-dimensional-looking propagator for a metric of ${\rm (2,2)}$ signature
\footnote{Notice that the fact that we obtain signature ${\rm (2,2)}$
is closely related with the choice of sign in the classical background
(\ref{2.4}).}
\begin{equation}
ds^{2}_{4d}=dx^{+}dq-dx^{-}dp.
\label{2.8}
\end{equation}

We will take $\Lambda\rightarrow 0$ for the remainder of the paper, 
defining a double-scaling limit where the level $K$ and the parameter 
$\epsilon$ are such that $g_{st}$ is finite and nonzero. 
The cubic vertex of order $g_{st}$ appearing in (\ref{2.7}) represents an 
$SO(2,2)$ Lorentz transformation of the cubic 
vertex $\Omega\{\partial_{+}\Omega ,\partial_{-}\Omega\}$ of (\ref{1.1}), whose
equation of motion is the Pleba\'nski equation (\ref{1.2}) 
\footnote{We note that the Pleba\'nski equation is not $SO(2,2)$ invariant, 
at least manifestly. For a discussion of the Lorentz symmetries of self-dual
gravity see \cite{PBR}.}. 
Its amplitudes \cite{PA} are closely related to  the ones of the closed ${\cal N}=2$ 
string \cite{OV}--\cite{L}.
However, we note that in the present model 
there is in addition an infinite series of higher point vertices.
At 
this point we can analytically continue in $x^{+},q,x^{-},p$ so that we get 
the flat K\"ahler metric 
\begin{equation}
ds^{2}_{4d}=dyd{\bar y}-dzd{\bar z}
\label{2.10}
\end{equation}
where $y,{\bar y},z,{\bar z}$ are complex coordinates corresponding to 
$x^{+},q,x^{-},p$ respectively. The field $\omega$ is then related to
deformations of the flat K\"ahler potential \cite{OV}.

In the limit when $\Lambda\rightarrow 0$ with $g_{st}$ kept fixed, the terms that remain 
in (\ref{2.7}) can be written more elegantly in terms of a group valued field 
$g=\exp ({\omega})$ as
\begin{equation}
S=\frac{1}{g_{st}^{2}}\int dx^{+}dx^{-} \T(qg\partial_{-}g^{-1}+pg\partial_{+}g^{-1})
\label{new.4}
\end{equation}
This is the Lagrangian that we would like to test as a field theory of closed ${\cal N}=2$ 
strings. 
The corresponding equation of motion is easily obtained by varying $g$ inside the trace 
and gives
\begin{equation}
\{g\partial_{-}g^{-1},q\}+\{g\partial_{+}g^{-1},p\}=0 \leftrightarrow 
\partial_{q}J_{+}-\partial_{p}J_{-}=0
\label{new.5}
\end{equation}
where $J_{\pm}=g\partial_{\pm}g^{-1}$. As before we also have the flatness condition
$\partial_{+}J_{-}-\partial_{-}J_{+}+\{J_{+},J_{-}\}=0$. If we solve (\ref{new.5}) by 
setting
$J_{-}=\partial_{q}\Omega$ and $J_{+}=\partial_{p}\Omega$ the flatness condition becomes 
$$
\partial_{+}\partial_{q}\Omega -\partial_{-}\partial_{p}\Omega +
\{\partial_{p}\Omega,\partial_{q}\Omega\}=0
$$
and in complex coordinate notation this is 
$$
\partial_{y}\partial_{\bar y}\Omega -\partial_{z}\partial_{\bar z}\Omega
-\{\partial_{\bar y}\Omega,\partial_{\bar z}\Omega\}=0
$$
which is the second form of the heavenly equation for self-dual gravity \cite{Pl,P}. 
In this way we formulate self-dual gravity in terms of the 
Lagrangian (\ref{new.4}) for a large $N$ group valued field $g$. Notice that expanding in 
$\omega$ where
$g=\exp(\omega)$ produces the complicated infinite series of terms that we already 
described above.
Moreover, the relation between the field $\Omega$ whose derivatives give the currents 
$J_{\pm}$ and the field
$\omega$ is highly non-linear and quite complicated. While such a field $\Omega$ could be 
described by a purely 
cubic action analogous to (\ref{1.1}) we would expect that action to be physically 
different, at the quantum level, from the more ``complete'' action (\ref{new.4}) 
for $g=\exp(\omega)$.


\section{A Reduction of Large ${\mathbf {\it N}}$ Self-Dual Yang-Mills}

As is well known, the self-dual Yang-Mills equations generate many known integrable 
systems by reduction.
If one takes the self-dual Yang-Mills equations of motion at large $N$,
that is if one takes the Lie algebra to be the Poisson algebra of functions on a surface 
$\Sigma$, one
obtains a 
six-dimensional equation of motion where as before the two extra variables come from the 
coordinates 
$q,p$ on $\Sigma$. In \cite{PP} it was shown that dimensionally reducing to four 
dimensions along two 
specific directions which ``mix'' space and ``color'' variables, produces self-dual 
gravity.
Along the same lines it is possible to show that the model in (\ref{2.7}) in the limit 
$\Lambda\rightarrow 0$
is a reduction of the large $N$ Donaldson-Nair-Schiff \cite{NaS} action for self-dual 
Yang-Mills theory.

Consider the self-dual Yang-Mills equations
\begin{equation}
F_{{\bar \mu}{\bar \nu}}=F_{\mu\nu}=0\;,\phantom{777}
\eta^{\mu {\bar \nu }}F_{\mu {\bar \nu }}=0,
\label{3.5}
\end{equation}
in the gauge where $A_{\bar \mu }=0$, $\mu,\nu =1,2$. The first equation in (\ref{3.5}) is 
solved by $A_{\mu }=g^{-1}\partial_{\mu }g$ so that the equation of motion,
Yang's equation, reads
\begin{equation}
\eta^{\mu {\bar \nu }}\partial_{\bar \nu }(g^{-1}\partial_{\mu }g)=0.
\label{3.6}
\end{equation}
We now consider the Donaldson-Nair-Schiff action \cite{NaS,LMNS}
\begin{equation}
S=\frac{i}{4\pi}\int d^{4}x\T (g^{-1}\partial^{\mu}gg^{-1}\partial_{\bar \mu}g)
+\frac{i}{12\pi}\int_{M_5} \omega\wedge\T(g^{-1}dg)^{3},
\label{3.7}
\end{equation}
where $\T$ denotes the trace on the Lie algebra of the gauge group, $G(N)$.
Parametrizing $g(x^{\mu},x^{\bar \mu})=\exp ({\hat \phi})$, one expands the 
action in powers of the field ${\hat \phi}$. In the limit $N\rightarrow\infty$,
we have a six-dimensional non-linear scalar field theory where the matrix field
${\hat \phi}(x^{\mu},x^{\bar \mu})$ becomes a scalar field 
$\phi(x^{\mu},x^{\bar \mu},q,p)$, where $q,p$ are the large $N$ color 
variables \cite{FIT}.
If we reduce by identifying $x^{\bar 1}$ with $q$ and $x^{\bar 2}$ with $p$,
that is if we impose
\begin{equation}
(\partial_{\bar 1}-\partial_{q})\phi = 
(\partial_{\bar 2}-\partial_{p})\phi =0,
\label{3.100}
\end{equation}
we obtain a four-dimensional theory. In the Leznov-Parkes gauge
this was seen to lead to the second heavenly equation of self-dual gravity 
\cite{PP}. In our case, we have a non-linear theory since the vertices follow 
from the four-dimensional WZW action evaluated in the large $N$ limit.
Expanding (\ref{3.7}), we have after some algebra
\begin{eqnarray}
\nonumber
S=\int d^{4}x\{\frac{1}{2}\phi\partial^{\mu}\partial_{\bar \mu}\phi
+\frac{1}{3!}\eta^{\mu{\bar \mu}}\varepsilon^{{\bar \rho}{\bar \nu}}
\phi\partial_{\mu}\partial_{{\bar \rho}}\phi\partial_{{\bar \mu}}
\partial_{{\bar \nu}}\phi
+\frac{1}{4!}\eta^{\mu {\bar \mu}}\varepsilon^{{\bar \lambda}{\bar \nu}}
\varepsilon^{{\bar \rho}{\bar \sigma}}\partial_{\mu}\partial_{{\bar \lambda}}
\phi
\partial_{{\bar \nu}}\phi\partial_{{\bar \mu}}\partial_{{\bar \rho}}\phi
\partial_{\sigma}\phi \\
+\frac{1}{5!}\varepsilon^{{\bar \rho}{\bar \sigma}}
\varepsilon^{{\bar \lambda}{\bar \nu}}\varepsilon^{{\bar \zeta}{\bar \xi}}
\partial_{{\bar \mu}}
\partial_{{\bar \rho}}\phi\partial_{{\bar \sigma}}\phi
\partial_{{\bar \zeta}}(\partial_{\mu}\partial_{{\bar \lambda}}
\phi\partial_{{\bar \nu}}\phi )\partial_{{\bar \xi}}\phi+\cdots\}
\phantom{7777}
\label{3.8}
\end{eqnarray}

These terms agree, to this order, with the vertices of the double scaling 
limit (\ref{new.4}) of the two-dimensional model (\ref{2.7}) as in (I). That is,
the Lagrangian (\ref{new.4}) also describes self-dual gravity as reduced large $N$ 
self-dual Yang-Mills theory.


\section{The Amplitudes}

In this section we will make some brief comments about the relation of the amplitudes 
in our model and the closed string amplitudes up to one-loop. The results have 
appeared in (I).
Since in our approach we generate an infinite series of higher point vertices,
we have to check that the contributions to the amplitudes coming from these vertices
are identical with the ${\cal N}=2$ closed string amplitudes. 
The fact that this infinite series of higher point vertices is present, and the 
close relation
of this theory with the WZW model, leads us to believe that it will 
have nice properties beyond the one-loop level. Moreover, the action 
(\ref{new.4}) provides a systematic expansion for determining these higher 
point vertices.
It is hoped that the fact that two of the four dimensions appear as
large $N$ color variables in disguise may shed some light into the problem
of matching string and field theory amplitudes at one-loop level
\cite{OV,BGI,M,L}. In fact, the large $N$ approach suggests a possible 
infra-red regulator for the momenta along ${\bar y},{\bar z}$. There is also 
the possibility that some specific choice of $\Sigma$, for example a torus 
$T^{2}$, will regulate these momenta even at infinite $N$.

The amplitudes will be $U(1,1)$ invariant functions of the components of the
external momenta $k_{i}$.
The metric is given by (\ref{2.10}) and the corresponding components of the 
momenta $k$ will be $k_{y},k_{{\bar y}},k_{z},k_{{\bar z}}$. The inner 
product will then be given by 
$k_{i}\cdot k_{j}=\overline{k_{j}\cdot k_{i}}=k_{iy}k_{j{\bar y}}-
k_{iz}k_{j{\bar z}}$.
Following \cite{OV,PA} we introduce the kinematical quantities
\footnote{The quantity ${\tilde c_{ij}}^{2}$ of \cite{OV} is given by
$a_{ij}{\bar a}_{ij}$. When $k_{i}$,$k_{j}$ and $(k_{i}+k_{j})$ are on-shell 
this becomes $c_{ij}^{2}$.},
\begin{eqnarray}
\nonumber
a_{ij}=-a_{ji}=k_{iy}k_{jz}-k_{jy}k_{iz}\;; \phantom{77777}
{\bar a}_{ij}=-{\bar a}_{ji}=
k_{i{\bar y}}k_{j{\bar z}}-k_{j{\bar y}}k_{i{\bar z}}\;;\\
k_{ij}=k_{i}\cdot k_{j}\;; \phantom{7777} s_{ij}=s_{ji}=k_{ij}+k_{ji}\;;
\phantom{7777} c_{ij}=-c_{ji}=k_{ij}-k_{ji}.
\label{3.1}
\end{eqnarray}

The propagator then becomes
\begin{equation}
\Delta (k,-k) = \frac{1}{k\cdot k} = \frac{1}{2s_{kk}}.
\label{3.45}
\end{equation}

The tree level 3-point function receives contributions only from the cubic 
vertex of (\ref{2.2}) and, on-shell, it is simply 
$V_{3}=g_{st}\, c_{13}{\bar a}_{13}$. In \cite{PA}, Parkes shows that this 
can be obtained from the usual cubic vertex in (\ref{1.1}), which is 
$a_{13}{\bar a}_{13}$, by an $SO(2,2)$ transformation. In our approach, since 
the ${\bar y}$ and 
${\bar z}$ coordinates are ``color'' variables, Lorentz transformations
are related to Lie algebra redefinitions of the field $\omega$, for example 
through commutators with $q$ and $p$. For example, if we were using exactly
the WZW model, the classical background would be a product of one 
anti-holomorphic term on the left and one holomorphic term on the right.
We could choose to define the quantum fluctuations either on the left or right,
or even in between these two terms. The resulting four-dimensional field
theory would have looked 
different, and the different quantum fields would be related by Lie algebra 
operations which would be connected with Lorentz transformations. Of course,
these quantum theories would be essentially equivalent. 

We could show directly that the tree level on shell 4-point function 
vanishes as a consequence of the identity $[{\bar a}_{12}{\bar a}_{13}s_{23}
+{\bar a}_{13}{\bar a}_{23}s_{12}
-{\bar a}_{12}{\bar a}_{23}s_{13}]=0$ valid on shell \cite{OV,PA}.
However, we can instead use the results of the last section where we have shown that our 
model is a dimensional reduction of large $N$ self-dual Yang-Mills theory.
 This property can  be used to conclude about the tree level on-shell amplitudes of our 
model from those of self-dual Yang-Mills theory. The later represents 
(at finite $N$) a field theory of open ${\cal N}=2$ strings with Chan-Paton factors
for the gauge group, say $U(N)$ \cite{M}. These amplitudes are given in a 
factorized sum over non-cyclic permutations 
$$
\sum_{\sigma}\T (T_{\sigma_{1}}T_{\sigma_{2}}\cdots T_{\sigma_{n}})
S_{n}(k_{1},...,k_{n}).
$$
In momentum space, the reduction (\ref{3.100}) is performed by identifying
the conjugate momenta $k_{q}=k_{{\bar 1}}$ and $k_{p}=k_{{\bar 2}}$. 
At tree level, by momentum conservation at the vertices, if these relations
are imposed on the external momenta they will be preserved throughout the 
Feynman graphs.
Then, the vanishing of the open ${\cal N}=2$ string amplitudes for $n\geq 4$, 
implies the vanishing of the corresponding $S_{n}$'s and also of the amplitudes for 
the model (\ref{2.2}),
indicating that it is indeed an appropriate field theory of the closed ${\cal N}=2$ 
string, at least at tree level.

At loop level however, there are integrations of the momenta along the loops,
and this argument does not apply. We therefore need a direct 
calculation.
In the pure cubic action for the field theory of the closed ${\cal N}=2$ string, 
the one-loop 3-point amplitude \cite{BGI} is less infra-red divergent than the
corresponding string amplitude, while it is also ultra-violet finite. 
In our case, with the action in (\ref{2.7}), we will obtain similar results.
However, as we mentioned before, the underlying $2+2$ structure, with two
dimensions coming from color, may if further explored solve this problem.

At one-loop, the 3-point amplitude may receive contributions from several 
types of graphs. This is to be contrasted with the cubic theory where only one
type of graph enters. Although, as we will see below, only this graph 
contributes also in our case, it is tempting to conjecture that the higher
point vertices in (\ref{2.7}) will be important to ensure good properties of 
the theory at more than one-loop level. Indeed, that is the case for the usual
two-dimensional WZW model. 

In our case, all the diagrams that could potentially give 
ultra-violet divergent contributions turn out to vanish due to various symmetries
of the integrands.
The only surviving term is the infra-red divergent one in the diagram 
containing just cubic vertices, which is proportional to
\begin{equation}
g_{st}^{3}(c_{13}{\bar a}_{13})^{3}\int_{0}^{\infty} d\alpha_{1}d\alpha_{2}
d\alpha_{3} \frac{\alpha_{1}^{2}\alpha_{2}^{2}\alpha_{3}^{2}}
{(\alpha_{1}+\alpha_{2}+\alpha_{3})^{8}}\exp(-(\alpha_{1}+\alpha_{2}+
\alpha_{3})\beta) (\int d^{4}p \exp(ip^{2})).
\label{3.10}
\end{equation}
where $\beta$ is a regulating parameter.
We stress that one should be careful in interpreting the integrals in
(\ref{3.10}), since it is not clear which regularizing prescription to use,
because of the peculiarities of the ${\rm (2,2)}$ signature. 

The Schwinger parameter integration in (\ref{3.10}) gives an infra-red 
divergence of the form 
$$
\int_{\varepsilon} ds \frac{1}{s^{3}}\sim\frac{1}{\varepsilon^{2}} 
$$
where $\varepsilon$ is some infra-red cut-off. This is indeed equivalent to 
the result of \cite{BGI} so that at one-loop this theory taken strictly at infinite $N$ 
gives the same result as the pseudo-dual cubic theory. However, one also has to note an 
infra-red divergence
associated with the singularity due to the ${\rm (2,2)}$ metric. In fact,
the gaussian momentum integral in (\ref{3.10}) is also divergent. In the 
present 
approach, two of the momentum components come from large $N$ color,
$(k_{q},k_{p})=(2\pi n_{q}/N, 2\pi n_{p}/N)$. This defines a natural 
infra-red regulator $\varepsilon =2\pi /N$. The transition from the sum over
$n_{q},n_{p}$ to the integral $\int dqdp$ then involves a factor of $N^{2}$
which is equivalent to $1/\varepsilon^{2}$. The total dependence on the 
infra-red regulator would then be 
$1/\varepsilon^{2}\cdot 1/\varepsilon^{2}=1/\varepsilon^{4}$ which would agree
with the ${\cal N}=2$ string one. This suggests that a further study of the finite $N$ 
theory could be relevant for solving the puzzle of matching string and field theory 
amplitudes.


\section{The Canonical Structure and the Current Algebra}

In this section we want to find the canonical structure defined by the Lagrangian 
(\ref{new.4}). We will obtain an infinite-dimensional current algebra of Poisson 
brackets similar to the current algebra of two-dimensional sigma models.
The action (\ref{new.4}) has a global symmetry given by $g \rightarrow gU$ where $U$ is 
an element of the group ``$SU(\infty)$''. The corresponding conserved currents are 
$Q=g^{-1}qg$ and $P=g^{-1}pg$ and obey the conservation law 
\begin{equation}
\partial_{+}P+\partial_{-}Q=0
\label{new.9}
\end{equation}
which is equivalent to the equation of motion (\ref{new.5}). The currents $Q$ and $P$
also satisfy a constraint $\{Q,P\}=1$ where we use the ad-invariance of the Lie-Poisson
bracket. The equation (\ref{new.9}) and the constraint can be directly related  
with those appearing in (2,2) self-dual gravity. Namely the Kahler form corresponding to 
the metric : 
$$
\Omega=\partial_{i} \partial_{\bar {j}} K dx^{i} \wedge dx^{\bar {j}} 
$$ 
where the $x^{i}$ are $y$ or $z$ 
($(x^{+},q,x^{-},p) \leftrightarrow (y,{\bar y},z,{\bar z})$) and $K$ is the Kahler 
potential can be written locally as:
\begin{equation}
\Omega= dP \wedge dz - dQ \wedge dy 
\label{form.1}
\end{equation}
where $P,Q$ are functions of $(y, \bar y, z, \bar z)$.
The hermiticity of the form will imply that the $ dy \wedge dz $ term should cancel 
giving equation (24). The Ricci flatness of the Kahler metric comes from the condition 
$\det(g_{i\bar {j}})=-1$, where $g_{i\bar {j}}=\partial_{i} \partial_{\bar {j}} K $ 
and from (\ref{form.1}) this is exactly the constraint we have above.

We now wish to find an Hamiltonian formulation for (\ref{new.4}) and the 
corresponding
Poisson brackets\footnote{We will denote these Poisson brackets for the field theory by 
$\{\;,\;\}_{PB}$
to avoid confusion with the $su(\infty)$ Lie-Poisson bracket $\{\;,\;\}$.}. 
The Poisson brackets of the theory can be obtained following Witten's method for finding 
the current algebra in the WZW model \cite{EW}. We choose light-cone quantization with 
$x^{+}$ playing the role of time. The only piece of the Lagrangian that contributes to 
the Poisson structure is the one containing derivatives in $x^{+}$ which when varied with 
respect to $g$ gives
\begin{equation}
\frac{1}{g_{st}^{2}}\int dx^{+}dx^{-}\T (\delta g\,g^{-1}\{p,\partial_{+}g g^{-1}\})
=\frac{1}{g_{st}^{2}}\int dx^{+}dx^{-} f_{abc}p^{a}(\delta g\,g^{-1})^{c}
(\partial_{+}g g^{-1})^{b}
\label{new.11}
\end{equation}
where $f_{abc}$ are the $sdiff(\Sigma)$ structure constants in some particular basis. The 
symplectic form will then be given by 
\begin{equation}
F_{ab}=\frac{1}{g_{st}^{2}}f_{abc}p^{c}
\label{new.15}
\end{equation}

where we raise and lower Lie algebra indices using the Killing metric as usual. The 
Poisson brackets of functions of $g$ will be given in terms of the inverse $F^{ab}$ of the 
matrix $F_{ab}$. In our case one just takes $\delta\Phi^{a}=(\delta g\, g^{-1})^{a}$
as coordinates of tangent vectors to the space of the fields $g$ 
and set \cite{EW}
\begin{equation}
\{ X,Y\}_{PB} =\sum_{a,b}F^{ab}\frac{\delta X}{\delta\Phi^{a}}\frac{\delta Y}
{\delta\Phi^{b}}.
\label{new.12}
\end{equation}
We assume that  $\Sigma$ is the plane\footnote{We expect this choice to be 
appropriate for the 
field theory of the uncompactified string. It would be interesting to discuss other 
possible 
choices for $\Sigma$ and to relate them to specific compactifications of the ${\cal N}=2$ 
string.} and take a basis of plane waves for $sdiff(\Sigma)$,
that is for the Poisson algebra of functions on $\Sigma$, with $L_{zw}=\exp(izq+iwp)$ 
where $z,w$ 
are real indices\footnote{Presumably to study the string compactified on $T^{2}$ we 
would take 
$\Sigma =T^{2}$ and would take only integer indices corresponding to periodic $q$ and 
$p$.} 
and $q,p$ are coordinates on $\Sigma$. 
The Poisson algebra becomes
\begin{equation}
\{L_{a},L_{b}\}=-(a\times b)L_{a+b}
\label{new.13}
\end{equation}
where the indices $a,b$ run over all pairs of real numbers and if $a=(z,w),b=(z',w')$ then
$(a\times b)=zw'-z'w$.
The Killing metric is given by $K_{ab}=\T(L_{a}L_{b})=(2\pi )^{2}\delta^{2} (a+b)$
and the structure constants are
$$
f_{abc}=-(2\pi)^{2} (a\times b) \delta^{2} (a+b+c)
$$
which is totally antisymmetric as usual. The components of $p$ are given by Fourier 
transforming 
$p$ with the result $p^{a}=p^{(z,w)} \equiv p(z,w)=i\delta (z)\delta^{'}(w)$. Then we 
have the Hamiltonian
\begin{equation}
H=-\frac{1}{g_{st}^{2}}\int dx^{-}\T (qg\partial_{-}g^{-1})
\label{new.14}
\end{equation}
and the symplectic form (\ref{new.15}) is given by 
\begin{equation}
F_{(z,w)(z',w')}=(2\pi)^{2}\frac{i}{g_{st}^{2}}(zw'-z'w) \delta (z+z') \delta^{'} (w+w')
\label{new.16}
\end{equation}
with inverse 
\begin{equation}
F^{(z,w)(z',w')}=-i\frac{g_{st}^{2}}{(2\pi)^{2}} \frac{1}{zw'-z'w} \delta(z+z')
\theta (w+w').
\label{new.17}
\end{equation}

We can now easily compute the Poisson brackets of the current $P$ with itself. We have 
the variation
\begin{equation}
\delta P_{(z,w)}=\delta\T (P \exp(izq+iwp))= g_{st}^{2}\delta\Phi^{a}F_{ab}
(g\cdot\exp(iqz+ipw)
\cdot g^{-1})^{b}
\label{new.18}
\end{equation}
and inserting $\delta P/\delta\Phi^{a}$ in the Poisson bracket (\ref{new.12}) 
$F_{ab}$ cancels against its inverse and we obtain at equal $x^{+}$
\begin{equation}
\{P_{(z,w)}(x^{-}),P_{(z',w')}(x^{-'})\}_{PB}=-g_{st}^{2}(zw'-z'w)P_{(z+z'.w+w')}
\delta (x^{-}-x^{-'}),
\label{new.19}
\end{equation}
which gives an infinite-dimensional current algebra that is the large $N$ limit of the 
current
algebra of the two-dimensional sigma model. In fact, we can also easily compute the Poison 
brackets 
$P$ with a function of the form $O=g^{-1}\cdot f(q,p)\cdot g$ since the same cancellation 
between 
$F_{ab}$ and $F^{ab}$ will occur with the result
\begin{equation}
\{P_{(z,w)}(x^{-}),O_{(z',w')}(x^{-'})\}_{PB}=-g_{st}^{2}(zw'-z'w)O_{(z+z',w+w')}
\delta (x^{-}-x^{-'}).
\label{new.20}
\end{equation}

We can also insert the basis elements $L_{zw}(q,p)$ in (\ref{new.20}) and obtain the 
Poisson 
algebra for the matrix elements of the current with the operators $O(x^{-},q,p)$,
\begin{eqnarray}
\nonumber
\{P(x^{-},q,p),O(x^{-'},q',p')\}_{PB}= \phantom{17171717171717171717171717}\\
\nonumber
=-g_{st}^{2}(\partial_{q}O\partial_{p}\delta (p-p')\delta(q-q')
-\partial_{p}O\partial_{q}\delta (q-q')\delta (p-p'))\;\delta(x^{-}-x^{-'})= \\
= -g_{st}^{2}\{O,\delta (q-q')\delta(p-p')\}\; \delta (x^{-}-x^{-'}).
\label{new.21}
\end{eqnarray}

We now introduce the functions $M=g^{-1}\cdot ((q^{2}+p^{2})/2)\cdot g$ and observe that
\begin{eqnarray}
\nonumber
\{Q,M\}=P \\
\{M,P\}=Q.
\label{new.22}
\end{eqnarray}
From this one has $M=1/2(Q^{2}+P^{2})$ and the action (\ref{new.4}) can be written
\begin{equation}
S=\int dx^{+}dx^{-}\; \frac{1}{2}\T ((Q^{2}+P^{2})(\partial_{+}Q-\partial_{-}P))
\label{new.23}
\end{equation}
which gives an Hamiltonian of the form
\begin{equation}
H=\int dx^{-} \T (\partial_{-}M \{M,Q\})
\label{new.24}
\end{equation}
and the Poisson structure
\begin{eqnarray}
\nonumber
\{Q(x^{-},q,p),M(x^{-'},q',p')\}_{PB}=\delta(q-q')\delta(p-p')\delta(x^{-}-x^{-'}) \\
\nonumber
\{Q(x^{-},q,p),Q(x^{-'},q',p')\}_{PB}=0 \\
\{M(x^{-},q,p),M(x^{-'},q'p')\}_{PB}=0
\label{new.25}
\end{eqnarray}
which together with the constraints (\ref{new.22}) does produce the equations of motion
\begin{equation}
\partial_{+}M = \{M,\partial_{-}M\} \;\;{\rm and}\;\; \partial_{+}Q = \{\partial_{-}Q,M\}
+2\{Q,\partial_{-}M\}
\label{new.26}
\end{equation}
which are equivalent to (\ref{new.5}).
In this way we obtain a canonical formulation for self-dual gravity which is similar
to the large $N$ limit of the canonical structure of the two-dimensional models.


\section{Conclusions}

In this paper we reviewed and further explored the formulation of the field theory of 
${\cal N}=2$ strings, $i.e.$ self-dual gravity, as a large $N$ two-dimensional 
sigma model, both at the classical and one-loop quantum levels.

We have seen that in contrast with the usual purely cubic action for the theory, we 
obtained a field theory for an $Sdiff(\Sigma)$ group valued field $g=\exp(\omega)$ 
which contains an infinite series of higher point vertices for the Lie algebra valued 
field $\omega$. At the 
quantum level, and in particular beyond one-loop, we expect this theory to have better 
properties 
than the cubic theory which is a pseudo-dual version \cite{FJ}. 
This model can also be obtained as a dimensional reduction of large $N$ self-dual 
Yang-Mills theory at 
large $N$ and as such it is closely related to a conjecture of Ooguri and Vafa \cite{OV}.

The mismatch between string and field theory one-loop amplitudes remains an
open problem although we 
believe and have given indications that a complete treatment of the large $N$ limit 
together with a  proper definition of the 
finite $N$ cutoff will help in solving this problem. As we mentioned in section 5, 
the structure of $4=2+2$ with two of the four dimensions coming from large $N$ color 
suggests 
by naive arguments that the carefully defined (in terms of powers of $N$) field theory 
amplitudes may coincide with the string amplitudes as desired.
It is exciting to speculate that the partonic structure \cite{Je} present in the large
$N$ formulation of self-dual gravity ultimately manages to improve its UV properties.

In the second part of the paper we found an Hamiltonian formulation of the theory which 
gives a close analog of the Hamiltonian structure of two-dimensional sigma models. 
This Poisson structure results in an infinite-dimensional 
algebra of currents and represents a large $N$ limit of the usual current algebras of 
two-dimensional sigma models. This construction gives an Hamiltonian approach to 
self-dual gravity which will be interesting to explore further in connection with 
matrix string theory, Hamiltonian and dimensional reductions and the study of Lorentz 
symmetries.

Other aspects of the problem that should be further studied are the inclusion of 
worldsheet $U(1)$ instantons in the amplitudes \cite{OV}-\cite{L}
and the consideration of the compactified string which we expect will lead us to 
more interesting choices for $\Sigma$.

Finally, we mention that the field theory of open ${\cal N}=2$ strings \cite{M} should 
in a similar way 
be very naturally expressable as a large $N$ two-dimensional sigma model. 
Indeed, Chan-Paton 
factors in the group $G$ can be incorporated by using an extension of 
$sdiff(\Sigma)$ by the 
Lie algebra of $G$ as in \cite{P}. We also expect that related considerations, 
but in a some 
more non-trivial way, extend this type of construction to the heterotic string theory 
of \cite{OV,KM}.


\end{document}